\documentclass[journal,twoside]{IEEEtran}

\usepackage{amsmath,amssymb}
\usepackage[T1]{fontenc}
\usepackage[dvips]{graphicx, color}
\usepackage{subfigure}
\usepackage{latexsym}
\definecolor{Bgreen}{rgb}{ .0, .55, .0 }
\definecolor{Red}{rgb}{ 1.0, .0, .0 }
\definecolor{Navy}{rgb}{ 0.0, .0, 1.0 }

\newcommand{\eqa}{\begin{eqnarray}}
\newcommand{\ena}{\end{eqnarray}}

\def\rn#1{\expandafter{\romannumeral#1}} % ���������[�}����

\newcounter{linenumber}
    {\end{list}}

\ifCLASSINFOpdf
\else
\fi
\begin{document}

\title{On the Performance of Exhaustive Search with Cooperating agents}

\author{\Large{Toni~Stojanovski$^1$,~and~Ljupco~Krstevski$^2$}
\\\normalsize{$^1$Faculty of Informatics, European University RM, Skopje, Macedonia, toni.stojanovski@eurm.edu.mk}
\\\normalsize{$^2$Faculty of Informatics, European University RM, Skopje, Macedonia, ljupco.krstevski@eurm.edu.mk}
\vspace{-0.5cm}
}
\maketitle
\pagestyle{empty}
\thispagestyle{empty}
\begin{abstract}
Despite the occurrence of elegant algorithms for solving complex problem, 
exhaustive search has retained its significance since many real-life problems exhibit no regular structure
and exhaustive search is the only possible solution.
The advent of high-performance computing either via multicore processors or distributed processors
emphasizes the possibility for exhaustive search by multiple search agents.
Here we analyse the performance of exhaustive search when it is conducted by multiple search agents.
Several strategies for cooperation between the search agents are evaluated.
We discover that the performance of the search improves with the increase in the level of cooperation.
Same search performance can be achieved with homogeneous and heterogeneous search agents
provided that the length of subregions allocated to individual search regions follow the differences in the speeds of heterogeneous search agents.
\end{abstract}

\begin{IEEEkeywords}Parallel algorithms, Search methods.
\end{IEEEkeywords}

\IEEEpeerreviewmaketitle

\section{Introduction}\label{intro}
\IEEEPARstart{E}xhaustive search consists of systematically enumerating all possible candidates for the solution and checking whether each candidate satisfies the problem's statement.
The number of candidate solutions to consider grows very rapidly with problem size, causing lengthy or even infeasible searches.

The continuing increase in computing power and memory sizes, and the advent of multicore processors and parallel and distributed programming \cite{Andrews-1999,CUDA-2010}
(combined effort of huge number of computers connected on the internet) increases the feasibility of exhaustive search and has revived interest in brute-force techniques for a good reason.
Many real-life problems reveal no regular structures to be exploited in the search for solutions, and this leaves exhaustive search as the only possible approach.
We can always try to design elegant and optimal algorithms in a quest for order of magnitude of performance improvements.
However, mathematical creativity is not guaranteed to give success for real-life problems.
As pointed out too many times in the past, it is more likely that order of magnitude improvements can be achieved due to program optimization, 
and to the clever use of limited computational resources in an exhaustive search.

There are many problems where the exhaustive search is the best solution. 
Radio SETI (Search for Extraterrestrial Intelligence) uses radio telescopes to listen for narrow-bandwidth radio signals from space, which are not known to occur naturally, so a detection would provide evidence of extraterrestrial technology.
Radio SETI has an insatiable appetite for computing power to cover greater frequency ranges with more sensitivity.
Early radio SETI projects have used special-purpose supercomputers, located at the telescope, to do the bulk of the data analysis.
SETI@home is doing radio SETI using a virtual supercomputer composed of large numbers of Internet-connected computers  \cite{setiathome2001}.
It was originally launched in May 1999.
As of 2008, five million people in 226 countries have participated volunteered their PCs to analyse data.
Combined, their PCs form Earth’s second most powerful supercomputer, averaging 482 TeraFLOPs and contributing over two million years of CPU time \cite{Siemion20101342}.

DES (Data Encryption Standard) \cite{Stallings-2005} is a secret-key block cipher that was selected by the National Bureau of Standards as an official Federal Information Processing Standard (FIPS) for the United States in 1976.
Its widespread international use gave rise to extensive cryptanalysis of DES.
None of the proposed cryptanalytic attacks was practically feasible, and eventually it was broken using exhaustive search of its 56-bit key region.
In 1998, Electronic Frontier Foundation \cite{Cocher-1999} built a custom-made machine named "Deep Crack" for the cost of US\$250,000,
and managed to decrypt the DES Challenge II-2 test message after 56 hours of work.
COPACOBANA machine built in 2006 \cite{copacobana} is the only other confirmed DES cracker, but unlike Deep Crack, COPACOBANA consist of commercially available, reconfigurable integrated circuits.
COPACOBANA's successor named RIVYERA reduced the time to recover a DES key to the average of a day \cite{rivyera}.

%TODO Example: Klima's collision attack on MD5 hash algorithm, 

%TODO Example: large number factorization (RSA challenges), 

%TODO Explain how the results in this paper are relevant to P2P applications [3]

In this paper we evaluate the performance of several strategies for exhaustive search depending on the following parameters:
\begin{itemize}
\item{Homogeneous or heterogeneous search agents according to their speed. What is the impact of the differences in speeds of search agents on the overall speed of the search?}
\item{Length of allocated search regions. What is the impact of the division of the search region on the speed of cooperative search?}
\end{itemize}

We will also demonstrate the optimum division of the search region.

Here is the overview of the paper. 
In Section~\ref{sec:ExhaustiveMethods} we present theoretical results on the average search time for exhaustive search.
Section~\ref{sect:homogeneous} we analyse exhaustive search with homogeneous cooperating agents, while in 
Section~\ref{sect:heterogeneous} we analyse cooperation of heterogeneous agents.
Section~\ref{sect:conclusion} concludes the paper and gives directions for future research.

\section{Exhaustive search methods} \label{sec:ExhaustiveMethods}
First we give a formal definition of exhaustive search.
Consider function $F: X\to [0,1]$ where $X$ is the discrete domain of $F$.
$X$ is also called the search region.
Assume that there is only one point $x_s\in X$ such that $F(x_s) =1$.
Otherwise, $F(x) = 0$ if $x\in X$ and $x\neq x_s$.
We call point $x_s$ solution of function $F$.
If function $F$ exhibits no regular structure, then finding the solution $x_s$ by exhaustively searching the domain $R$ can be the only option.
Exhaustive search attempts to find the point $x_s$ by repeatedly calculating $F(x)$ for all points $X\in R$ until $x_s$ is found such that $F(x_s) =1$.
We assume that the region is cooperatively searched by $m$ agents $a_1, a_2, \dots , a_m$.
We are interested in the impact of the cooperation between multiple search agents on the performance of the search.
Search performance is measured bu the average time required to find the only solution $x_s$.

\begin{figure}[ht]
\begin{center}
\includegraphics[width=76mm]{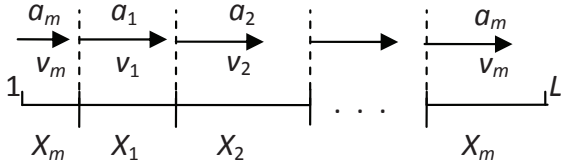}
\end{center}
\caption{Exhaustive search with multiple search agents.}
\label{fig-OneDirection}
\end{figure}

For sake of simplicity and clarity, we analyse exhaustive search of one-dimensional region $X = [1, L]$.
However, results are equaly valid for multi-dimensional regions since they can be easily converted into a one-dimensional region.
Each agent $a_i$ is allocated a subregion $X_i \subset X$ which it searches with speed $v_i$.
An obvious requirement is that $ \bigcup_{i=1}^{i=m}X_i= X$.
Each agent starts its search from the starting point of the allocated subregion.
If an agent reaches the end of the search region, then it continues from the beginning of the search region.
If $m$-th agent reaches point $L$, then it continues the search from point 1.
This is illustrated in Fig.~\ref{fig-OneDirection}

The process of allocation of search subregions to search agents depends on the type of communication between search agents and the central server, 
whether the central server knows the number $m$ of search agents and the speed of each agent, 
and whether the number of search agents is determined before the search starts or additional agents can join the search at a later time.
One possibility is that the central server can allocate a subregion to each search agent depending on the number $m$ of search agents and the speed of each agent.
Another possibility is that each agent chooses its own starting point for the search irrespective of the starting points of the other search agents.
Search continues until a solution is found, or a "stop" command is received from the central server.

We assume uniform probability distribution of the solution in the search region $X$, that is, each point in the search region $X$ is equally probable to be the solution $x_s$.
Average search time is calculated using the following formula
\begin{equation}\label{eq1i}
E(t) = \sum_l\sum_v p_v(v)\Pr\{x_s\in R_l\}\frac{l}{2v}
\end{equation}
where (i) $\Pr\{x_s\in R_l\}$  is the probability that the solution is in region $R_l$ with length $l$;
(ii) $p_v(v)$ is the probability that the speed of a search agent is $v$; 
and (iii) $l/(2v)$ is the average time to find the solution in a region of length $l$ by agent with speed $v$.

In Eq.~(\ref{eq1i}) it is assumed that the two random variables are mutually independent. 
If not, then in Eq.~(\ref{eq1i}) one needs to use joint probability distribution function
\begin{equation}\label{eq1j}
E(t) = \sum_l\sum_v \Pr\{v, x_s\in R_l\}\frac{l}{2v}
\end{equation}

If the search region with length $L$ is divided into $m$ subregions, and length $l$ of subregions is a random variable with probability distribution function $p_l(l)$, 
then one can easily show that
\begin{equation}\label{eqPrSinRl}
\Pr\{x_s\in R_l\} = \frac{mp_l(l)l}{L}
\end{equation}
where $mp_l(l)$ is the number of regions with length $l$, $mp_l(l)l$ is the total number of points belonging to regions with length $l$, 
and the ratio $mp_l(l)l/L$ is the probability that the solution $x_s$ is in a region with length $l$.
Substituting Eq.~(\ref{eqPrSinRl}) in Eq.~(\ref{eq1j}), one obtains
\begin{equation}\label{eq2j}
E(t) = \frac{m}{2L}\sum_l\sum_v p_{v,l}(v, l)\frac{l^2}{v}
\end{equation}
If $v$ and $l$ are mutually independent random variables, then Eq.~(\ref{eq2j}) transform into 
\begin{equation}\label{eq2i}
E(t) = \frac{m}{2L}\sum_l\sum_v p_v(v)p_l(l)\frac{l^2}{v} = \frac{m}{2L}E(\frac{1}{v})E(l^2)
\end{equation}

\section{Exhaustive search with homogeneous agents}\label{sect:homogeneous} 
In this section we additionally assume that all search agents are homogeneous, that is, 
all search agents are with same speed $V$ and 
$p_v(v) = \begin{cases}
1 &\text{for } v = V\\
0 &\text{otherwise}
\end{cases}$.
Here we compare three methods for division of the search region, which are explained next.

\subsection{Equal subregions}
%
%All $m$ search subregions are of equal size. 
This can occur if the number of search agents $m$ is known in advance, and two directional communications exists between central server and search agents.
Thus a search subregion is allocated and communicated to each search agent in advance.
In this case 
\begin{equation}\label{eq2l}
p_l(l) = \begin{cases}
1 &\text{for } l = \frac{L}{m}\\
0 &\text{otherwise}
\end{cases}
\end{equation}
Thus Eq.~(\ref{eq2i}) yields
\begin{equation}\label{eq2t}
E(t) = \frac{L}{2m}E(\frac{1}{v}) = \frac{L}{2mV}
\end{equation}

\subsection{Semi-equal subregions}
Search subregion is allocated and communicated to each search agent by the central server as it registers.
Then search subregion allocated to a newly registered agent depends on the current number of search agents.
This case can occur if the number of search agents is not known in advance, and two directional communications exists between central server and search agents.
New search agents can register at run time.
Let $[1, L]$ denote the entire search region.
Then the first agent will start its search from point 1.
If there is only one search agent in the system, then it searches the whole range.
If there is a second search agent in the system, then its searching subregion is the second half of the whole region, that is, the second agent will start the search from point $1+L/2$. 
If the searched solution is located in the second half of the range, then the second search agent will find it faster than the first search agent.
Third agent will start from point $1+L/4$, and the next agent will start from point $1+3L/4$.
The next agents will start the search from points $1+L/8$, $1+3L/8$, $1+5L/8$, $1+7L/8$ etc.
Thus, the searching subregions are shrinking as the number of search agents grows.
Each time a new search agent joins the search, it is given to search the second half of the currently largest search subregion. 
In the semi-equal subregions method, for $m = 2^0, 2^1, 2^2, 2^3,\dots$,  $p_l(l)$ and $E(t)$ are again given by Eq.~(\ref{eq2l}) and Eq.~(\ref{eq2t}). Hence the semi-equal subregions method is identical to the equal subregions method. 
Otherwise, for $2^n < m < 2^{n+1}$, it is easy to show that the probability that a search subregion is with length $l$ is given by
\begin{equation}\label{eq3}
p_l(l) = \begin{cases}
\frac{2^{n+1}-m}{m} &\text{for } l = \frac{L}{2^n}\\
\frac{2m- 2^{n+1}}{m} &\text{for } l = \frac{L}{2^{n+1}}\\
0 &\text{otherwise}
\end{cases}
\end{equation} 
The average search time calculated using Eq.~(\ref{eq3}) is higher than the average search time for the {\em equal subregions} method obtained from Eq.~(\ref{eq2t}).

\subsection{Random subregions}
Each search agent starts from a randomly chosen starting point, thus randomly choosing its search subregion.
Therefore, the size of the subregion searched by an agent can vary between 0 and the size $L$ of the entire search region.
This kind of search can occur if, for example, one directional communication exists from search agents to central server.
Number of search agents is not known in advance, and new search agents can join at run time.
The one directional communication is used by a search agent to communicate to the central server when the solution is found.

\begin{figure}[ht]
\begin{center}
\includegraphics[width=90mm]{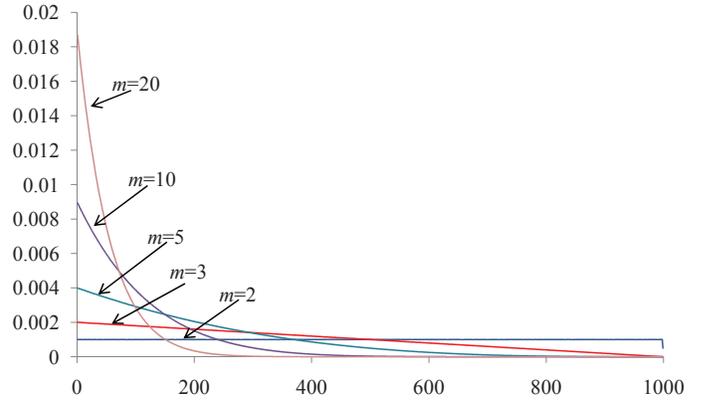}
\end{center}
\caption{Numerically calculated probability function $p_l(l)$ for $L = 1000$ and $m = 2, 5, 10, 20, 30$.}
\label{fig-ProbFunc}
\end{figure}

For given $L$, $m$ and when agents randomly choose starting points independently from each other, it is possible to analytically calculate $p_l(l)$.
However, our main interest in this paper is the average search time and comparison with the other two methods {\em Equal subregions} and {\em Semi-equal subregions}.
Therefore, we have taken the simpler approach and have numerically calculated the probability function $p_l(l)$ for $L = 1000$ and $m = 2, 3, 4,\dots, 32$ using the Monte-Carlo method \cite{Metropolis-1953}.
Figure~\ref{fig-ProbFunc} depicts the calculated probability function $p_l(l)$.
For $m=2$, $p_l(l)$ is uniformly distributed probability function in the region $[0, 999]$.
For $m=3$, $p_l(l)$ is linearly decreasing function in the region $[0, 999]$.
For $m>3$, $p_l(l)$ is monotonically decreasing polynomial function in the region $[0, 999]$.

\subsection{Comparison}
\begin{figure}[ht]
\begin{center}
\includegraphics[width=90mm]{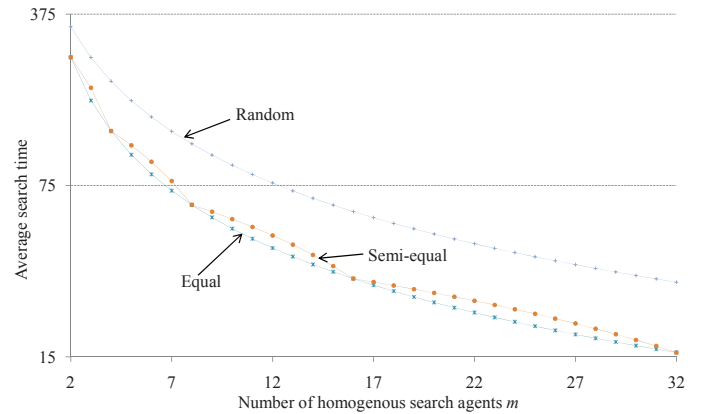}
\end{center}
\caption{Comparison of performances of the three searching methods.}
\label{Homogenous3}
\end{figure}

Figure~\ref{Homogenous3} gives the average search time (y-axis) for the three methods depending on the number $m$ of homogeneous search agents (x-axis).

As expected {\it Equal subregions} method produces the best performance, i.e., the shortest average search time.
{\it Semi-equal subregions} method produces performances which are close to the ones produced by the {\it Equal subregions} method.
{\it Random subregions} method results in significantly higher average search times.
For example, 10 search agents using the {\it equal subregions} method will produce same performance as 19 search agents using the {\it random subregions} method,
and 16 search agents using the {\it equal subregions} method will produce same performance as 31 search agents using the {\it random subregions} method.

\section{Exhaustive search with heterogeneous agents}\label{sect:heterogeneous} 
In this section we consider the performance of exhaustive search when the cooperating agents are heterogeneous, that is, agents search the search region $X$ with different speed.
Average search time is calculated using Eq.~(\ref{eq2j}) and Eq.~(\ref{eq2i}).
An interesting consequence from Eq.~(\ref{eq2i}) is following: 
if $v$ and $l$ are mutually independent random variables, then average search time depends on $E(1/v)$.
In other words, heterogeneous agents will provide the same search performance as homogeneous agents whose search speed is $V = E(1/v)$.
In our experiments, searching agents have speed which is a random variable with the following probability function: $P_v(0.5) = 0.3$, $P_v(1.0)=0.3$ and $P_v(1.375) = 0.4$.
One can easily calculate $E(1/v) = 1.19091$.

We analyse the performance of exhaustive search with cooperating heterogeneous agents for the following three search strategies.

\subsection{Strategy 1: One-directional search}
Each agent is randomly allocated a starting point for the search.
Then the agent searches the region from its starting point untill the starting point for the next agent.
Starting point is a uniformly distributed random variable in the region $[1, L]$.
%Each agent is randomly allocated a search subregion, that is, a starting point for the search is randomly assigned to each agent.
Length of allocated search subregions is a random variable whose probability function is depicted in Fig.~\ref{fig-ProbFunc}.
Then each agent is searching its own subregion (see Fig.~\ref{fig-OneDirection}).
It is possible that agent $a_i$ finishes the search of its region $X_i$ before the solution $x_s$ is found by any of the agents.
Then agent $a_i$ continues with the search of the subregion $X_{i+1}$ until a "stop" command is received by the central server.
Let $v_{min}$ and $v_{max}$ denote the minimum and maximum speed of search agents, and $l_{min}$ and $l_{max}$ denote the minimum and maximum length of search subregions.
Then provided that 
\begin{equation}
\frac{v_{max}}{v_{min}} < \frac{l_{min}+l_{max}}{l_{max}}
\end{equation}
agent $a_{i+1}$ will finish searching subregion $X_{i+1}$ before agent $a_i$ will finish searching both regions $X_i$ and $X_{i+1}$ for all $i=1,2,\dots , m$.
In other words, each agent is responsible for the search of its allocated subregion and will receive no help from other agents in the search of its subregion.

\subsection{Strategy 2: Two-directional search}
\begin{figure}[ht]
\begin{center}
\includegraphics[width=76mm]{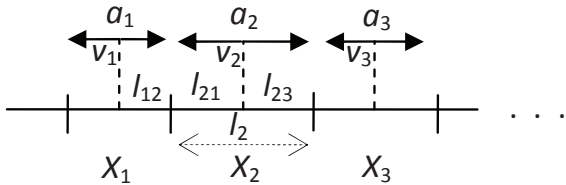}
\end{center}
\caption{Exhaustive search with cooperative agents in two directions.}
\label{fig-Helping}
\end{figure}

Similar to the previous case, each agent is randomly allocated a search subregion by means of an assigned starting point.
However, neighbouring agents can help each other in the following manner:
each agent $a_i$ conducts the search of its subregion $X_i$ in two directions: to the left and to the right from the assigned starting point (see Fig.~\ref{fig-Helping}).
If the speed of search for agent $a_i$ is $v_i$, then the search to the left side is conducted with speed $v_1/2$. Same speed applies for the search to the right side.
Searching in both directions improves the cooperation between neighbouring agents.
Assume that the distance between the starting points for agents $a_1$ and $a_2$ is $d_{1,2}$, and 
the distance between the starting points for agents $a_2$ and $a_3$ is $d_{2,3}$.
If $a_2$ is faster than $a_1$, then $a_2$ will search larger portion of $d_{1,2}$ than $a_1$.
Similarly, if $a_2$ is slower than $a_3$, then $a_2$ will search smaller portion of $d_{2,3}$ than $a_3$.
In other words, agents help their slower neighbours, and get help from faster neighbours.
It is rather straightforward to show that, for example, agent $a_2$ searches a subregion with length 
$l_2 = l_{21}+l_{23}$
where $l_{21}=\frac{l_2+l_1}{v_2+v_1}v_2$ and $l_{23}=\frac{l_2+l_3}{v_2+v_3}v_2$.
Time spent by agent $a_2$ to search the subregion $X_2$ is given by the following equation:
$t_2=(l_{21}+l_{23})/v_2$.

\begin{figure}[ht]
\begin{center}
\includegraphics[width=90mm]{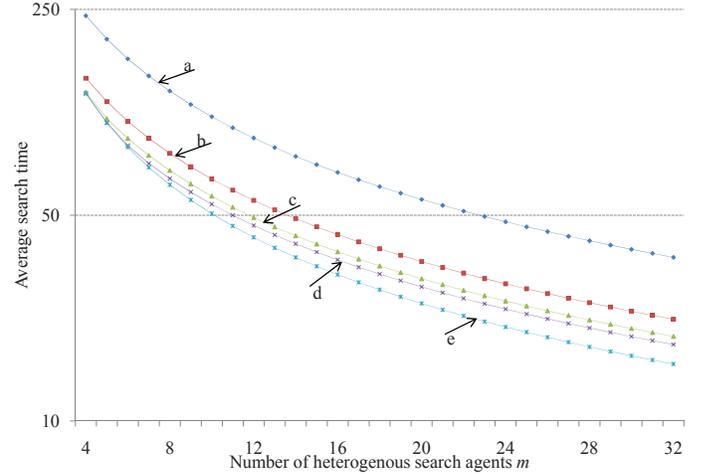}
\end{center}
\caption{Average search time for cooperating heterogeneous search agents. a) random; b) joint search  in groups of two agents; 
c) joint search  in groups of three agents;
d) joint search  in groups of four agents;
e) length of subregion is determined by agent's speed.}
\label{fig-ComparisonHeterogenousThree}
\end{figure}

Figure~\ref{fig-ComparisonHeterogenousThree} gives the average search time (y-axis) for the three methods depending on the number of heterogeneous search agents (x-axis).
It is obvious that the search strategy with two neighbours helping each other (two-directional search) significantly reduces the search time compared to the one-directional search strategy where the neighbouring agents do not help each other.
An obvious question arises: can the search time be further reduced if agents help each other in groups of three, that is, agents in groups of three jointly search an allocated subregion?
Defining the strategy for cooperation amongst groups of three neighbours and for joint search of an allocated subregion is beyond the scope of this paper.
Here we are interested only in the search performance.
Figure~\ref{fig-ComparisonHeterogenousThree} confirms our intuitive expectations: joint search of allocated regions in groups of three agents further reduces the average search time.
Further increasing the number of agents that jointly search a subregion to four additionally reduces the search time, as shown in Fig.~\ref{fig-ComparisonHeterogenousThree} too.

\begin{figure}[ht]
\begin{center}
\includegraphics[width=56mm]{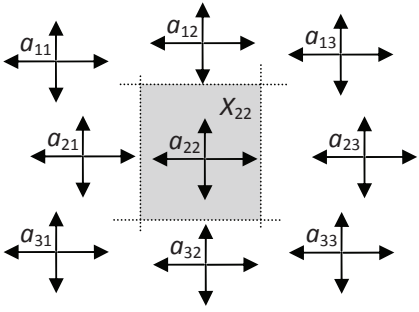}
\end{center}
\caption{Exhaustive search in a two-dimensional search region and cooperation with four neighbours.}
\label{fig-4Helping}
\end{figure}

Figure~\ref{fig-4Helping} gives a possible strategy for joint search by groups of four search agents in a two-dimensional search space $X$.
Each agent should search in four directions simultaneously, that is, in two directions for each of the two dimensions.
If the speed of agent $a_i$ is $v_i$, then the search speed in each of the four directions will be $v_i/4$.
If a subregion $X_{1,2,3,4}$ is jointly searched by four agents $a_1, a_2, a_3, a_4$ with search speeds $v_1, v_2, v_3, v_4$, respectively,
then agent $a_1$ will search $v_1/(v_1+v_2+v_3+v_4)$ parts from region $X_{1,2,3,4}$,
agent $a_2$ will search $v_2/(v_1+v_2+v_3+v_4)$ parts from region $X_{1,2,3,4}$ etc.
Subregion $X_{22}$ searched by agent $a_{22}$ is shown as a shaded square in Fig.~\ref{fig-4Helping}.
Careful examination of the boundaries of the subregion $X_{22}$ reveals that 
$v_{22}<v_{23}, v_{22}>v_{32}, v_{22}<v_{12}, v_{22}=v_{21}$.

Figure~\ref{fig-4Helping} also gives a hint on a possible strategy for cooperation between $2^N$ neighbouring agents, for $N=2,3,\dots$.
If the search region is $N$-dimensional, then each agent can be assigned a random starting point and the agent can start the search in $2N$ directions - two directions for each dimension.
If the dimension of the search region $X$ is less than $N$, e.g. 1, then one can transform $X$ into a $N$-dimensional search region.

\subsection{Strategy 3: Subregion's length is proportional to agent's speed}
Each search agent is allocated search subregion whose length is proportional to agent's speed. 
Faster agents get larger subregions; slower agents get smaller subregions.
The size $l_i$ of the subregion $X_i$ allocated to agent $a_i$ is determined by the following equation:
\begin{equation}\label{eq5}
l_i=\frac{v_i L}{v_1+v_2+\ldots+v_m}
\end{equation}
Each agent will finish the search of its allocated subregion at the same time 
\begin{equation}
t_i = \frac{l_i}{v_i}=\frac{L}{v_1+v_2+\ldots+v_m}
\end{equation}
and the average search time is given by
\begin{equation}
E(t) = \frac{L}{2mE(v)}
\end{equation}
This strategy is the optimum strategy for division of the search region for heterogeneous agents.

\subsection{Comparison}
As we see from Fig.~\ref{fig-ComparisonHeterogenousThree}, in the case of heterogeneous agents,
average search time decreases and the search performance improves as the cooperation between agents grows.
One-directional search is the worst search strategy: cooperation is reduced to the division of the search region between the search agents.
If faster agents are enabled to help slower agents e.g. by searching the allocated one-dimensional subregion in two directions, then the average search time reduces dramatically.
If the number of agents $n$ that jointly search a subregion grows, then the averages search time further reduces.
For example, one-directional search with $m=23$ agents,
two-directional search with $m=14$ agents for groups of two agents,   
two-directional search with $m=12$ agents for groups of three agents, 
two-directional search with $m=11$ agents for groups of four agents, 
and optimum search with $m=10$ agents produce simillar average search times.
For $n=m$, two-directional search strategy converges to the optimum strategy and each agent searches a subregion whose size is proportional to the agent's speed.

We also note that the optimum strategy ({\em Subregion's length is proportional to agent's speed}) for heterogeneous agents produces 
same search performance as the {\it Equal subregions} method for homogeneous agents,
if $E(v)$ for the heterogeneous agents is equal to the search speed $V$ of homogeneous agents.

\section{Conclusion} \label{sect:conclusion}
We have analysed the performance of exhaustive search by cooperative search agents.
Both homogeneous and heterogeneous agents are analysed.
Performance of exhaustive search by cooperative search agents improves and average search time decreases as the level of cooperation increases.
Optimum performance is achieved if the central server knows the number and speed of search agents,
and then each agent is allocated a search subregion with length proportional to agent's speed.
If the search region is with high-dimension $N$ close to the number of agents $m$ and two-directional search strategy is employed,
then search performances close to the optimum can be achieved.
Results given in this paper are presented for exhaustive search, but they are equally valid for other search methods where multiple search agents cooperate.

%%%===========================  Appendix =================
\ifCLASSOPTIONcaptionsoff
  \newpage
\fi

\bibliographystyle{IEEEtran}
\bibliography{ref_bfs}

\begin{thebibliography}{1}
\providecommand{\url}[1]{#1}
\csname url@samestyle\endcsname
\providecommand{\newblock}{\relax}
\providecommand{\bibinfo}[2]{#2}
\providecommand{\BIBentrySTDinterwordspacing}{\spaceskip=0pt\relax}
\providecommand{\BIBentryALTinterwordstretchfactor}{4}
\providecommand{\BIBentryALTinterwordspacing}{\spaceskip=\fontdimen2\font plus
\BIBentryALTinterwordstretchfactor\fontdimen3\font minus
  \fontdimen4\font\relax}
\providecommand{\BIBforeignlanguage}[2]{{%
\expandafter\ifx\csname l@#1\endcsname\relax
\typeout{** WARNING: IEEEtran.bst: No hyphenation pattern has been}%
\typeout{** loaded for the language `#1'. Using the pattern for}%
\typeout{** the default language instead.}%
\else
\language=\csname l@#1\endcsname
\fi
#2}}
\providecommand{\BIBdecl}{\relax}
\BIBdecl

\bibitem{Andrews-1999}
G.~R. Andrews, \emph{Foundations of Multithreaded, Parallel, and Distributed
  Programming}.\hskip 1em plus 0.5em minus 0.4em\relax Addison Wesley, 1999.

\bibitem{CUDA-2010}
J.~Sanders and E.~Kandrot, \emph{CUDA by example : an introduction to
  general-purpose GPU programming}.\hskip 1em plus 0.5em minus 0.4em\relax
  Addison Wesley, 2010.

\bibitem{setiathome2001}
E.~Korpela, D.~Werthimer, D.~Anderson, J.~Cobb, and M.~Lebofsky,
  ``Seti@home--massively distributed computing for seti,'' \emph{Computing in
  Science and Engineering}, vol.~3, no.~1, pp. 78--83, 2011.

\bibitem{Siemion20101342}
\BIBentryALTinterwordspacing
A.~Siemion, J.~V. Korff, P.~McMahon, E.~Korpela, D.~Werthimer, D.~Anderson,
  G.~Bower, J.~Cobb, G.~Foster, M.~Lebofsky, J.~van Leeuwen, and M.~Wagner,
  ``New seti sky surveys for radio pulses,'' \emph{Acta Astronautica}, vol.~67,
  no. 11-12, pp. 1342 -- 1349, 2010, special Issue on Searching for Life
  Signatures. [Online]. Available:
  \url{http://www.sciencedirect.com/science/article/B6V1N-4YDR2C1-1/2/22626e9907c44b5529edf5e30623f123}
\BIBentrySTDinterwordspacing

\bibitem{Stallings-2005}
W.~Stallings, \emph{Cryptography and Network Security, Principles and
  Practices}.\hskip 1em plus 0.5em minus 0.4em\relax Prentice Hall, 2005.

\bibitem{Cocher-1999}
P.~C. Cocher, ``Breaking des,'' \emph{CryptoBytes}, vol.~4, no.~2, pp. 1--5,
  1999.

\bibitem{copacobana}
\BIBentryALTinterwordspacing
\emph{COPACOBANA: A Codebreaker for DES and other Ciphers}.\hskip 1em plus
  0.5em minus 0.4em\relax SciEngines. [Online]. Available:
  \url{http://www.copacobana.org/}
\BIBentrySTDinterwordspacing

\bibitem{rivyera}
\BIBentryALTinterwordspacing
\emph{RIVYERA S3-5000}.\hskip 1em plus 0.5em minus 0.4em\relax SciEngines.
  [Online]. Available:
  \url{http://www.sciengines.com/products/computers-and-clusters/rivyera-s3-5000.html}
\BIBentrySTDinterwordspacing

\bibitem{Metropolis-1953}
N.~Metropolis, A.~W. Rosenezluth, M.~N. Rosenbluth, A.~H. Teller, and
  E.~Teller, ``Equation of state calculations by fast computing machines,''
  \emph{J. Chem. Phys.}, vol.~21, p. 1087, 1953.

\end{thebibliography}
\end{document}